\documentstyle [12pt,epsf] {article}
\begin{document}
\bibliographystyle{unsrt}

\pagestyle{empty}               
	
\rightline{\vbox{
	\halign{&#\hfil\cr
	&RHIC Spin Note\cr}}}

\rightline{\vbox{
	\halign{&#\hfil\cr
	&January 2002\cr}}}
\vskip 1in
\begin{center}
{\Large\bf
{Energy dependence of CNI analyzing power for proton-carbon scattering}}
\vskip .5in
\normalsize
T.L.\ Trueman \footnote{This manuscript has been authored
under contract number DE-AC02-76CH00016 with the U.S. Department
of Energy.  Accordingly, the
U.S. Government retains a non-exclusive, royalty-free license to
publish or reproduce the published form of this contribution, or
allow others to do so, for U.S. Government purposes.}\\
{\sl Physics Department, Brookhaven National 
Laboratory, Upton, NY 11973}
\end{center}
\vskip 0.5
in
\begin{abstract} We use a simple Regge model to determine the energy
dependence of the analyzing power for $pC$ scattering in the CNI region. We
take the model of Cudell et al which determines the Regge couplings and
intercepts for the $I=0$, non-flip Regge exchanges (Pomeron, $f1$ and
$\omega$) and extend it to the spin-flip amplitudes by allowing each of
these exchanges to have independent spin-flip factors $\tau_P, \tau_f $ and
$\tau_\omega$. Using this we show that by making measurments at two separate
energies, with polarization known at one energy, one can fix the ratios of the
analyzing power at any energy. By
making an additional assumption that is reasonable, but not necessarily
true, namely
$\tau_\omega=\tau_f$, we show that one can predict the energy dependence of
the analyzing power {\em using the existing E950 data}. We present the
corresponding predictions for beam energies of 100 GeV and 250 GeV protons
on a fixed carbon target based on a fit to the Spin 2000 data. Finally, we
discuss the relation of these results to the $pp$ CNI analyzing power. 
\end{abstract}
\vfill \eject \pagestyle{plain}
\setcounter{page}{1}
\section{} We begin with the parametrization of $pp$ elastic scattering given
by Cudell et al \cite{Cudell}, though one might do the same thing using other
parametrizations such as that of Block et al \cite{Block}. Since it is
known that the elastic, non-flip scattering is overwhelmingly $I=0$
exchange, even at 24 GeV/c \cite{np data}, we will assume the Regge
couplings that they determine are for the $I=0$ families and so directly
applicable to $pC$ scattering. The form they assume for the forward
amplitude then has the form

\begin{equation}
g_0(s,0)= g_P(s) + g_f(s) + g_\omega (s)
\end{equation}
with
\begin{eqnarray}
g_P(s)&=&-X s^\epsilon (\cot\!\frac{\pi}{2}(1 + \epsilon) - i),\\
\newline g_f(s)&=&-Y s^{-\eta} (\cot\!\frac{\pi}{2}(1 - \eta) - i), \\
\newline g_\omega(s)&=&- Y's^{-\eta'} (\tan\!\frac{\pi}{2}(1 - \eta') + i) 
\end{eqnarray}
normalized that $Im(g0(s))= \sigma_{\rm tot}(s)$. The values of the
parameters given by them are

\begin{eqnarray}
 \epsilon= 0.0933,\, \eta =0.357, \,\eta' =0.560 \\ \newline
 X=18.79,\, Y=63.0,\, Y'=36.2.
\end{eqnarray}

Our model is that the spin-flip $pp$ $I=0$ exchange
amplitude $g_5(s,t)$ is given by

\begin{eqnarray} 
g_5(s,t) &=& \tau(s) \frac{\sqrt{-t}}{m} g_0(s,t) \\
{} &=& \frac{\sqrt{-t}}{m} \{\tau_P\, g_P(s) + \tau_f\, g_f(s) +
\tau_\omega\, g_\omega(s)\}.
\end{eqnarray}
where $\tau(s)$ depends on energy but not on $t$ over the CNI range. It is in
general neither real nor constant in $s$ and is given by

\begin{equation}
\tau(s) = \{\tau_P\, g_P(s) + \tau_f\, g_f(s) + \tau_\omega\,
g_\omega(s)\}/g_0(s,0)
\end{equation}
 where the $\tau_i$'s are energy-independent, {\em real} constants. The
phases of the amplitudes come only from the energy dependence as given in
Eq.(1). This is the key assumption from Regge theory which we need: as a
result the real and imaginary parts of $\tau(s)$ are given at each energy
in terms of the three real constants $\tau_P, \tau_f$ and $\tau_\omega$.

In a recent paper \cite{K&T} it was shown under rather general assumptions
that the spin-flip factor for proton-nucleus scattering $\tau_{pA}(s)$ is
equal to the $I=0$ part of the proton-proton spin-flip factor $\tau(s)$. From
here on we will use this result to study the energy dependence of the $pC$
analyzing power, and will return to the question of $pp$ analyzing power at
the end of the note.

To determine the three real parameters $\tau_P, \tau_f$ and $\tau_\omega$ we
need three equations. At each energy, the fit to the small $t$ behaviour
determines two quantities, $P(s)(\kappa/2 -Re[\tau(s)])$ and
$P(s)Im[\tau(s)]$. Thus if we know the polarization at one energy, $s_0$,
then we have two of the needed equations:
\begin{eqnarray}
Re[\tau(s)]&=& \tau_P\, Re[g_P(s)/g_0(s)] + \tau_f\,
Re[g_f(s)/g_0(s)] + \tau_\omega \,Re[g_\omega(s)/g_0(s)], \\
Im[\tau(s)]&=& \tau_P\, Im[g_P(s)/g_0(s)] + \tau_f \,Im[g_f(s)/g_0(s)]
+\tau_\omega \,Im[g_\omega(s)/g_0(s)],
\end{eqnarray}
evaluated at $s=s_0$.
If we measure the asymmetry but not  the polarization at some other 
energy $s$ then all we can obtain is the shape of the curve characterized by
the energy-dependent parameter
$S(s)$
\begin{equation}
S(s)=\frac{Im[\tau(s)]}{\kappa/2 - Re[\tau(s)]}.
\end{equation}
Given a measured value $S(s_1),\, s_1 \neq s_0$ we can use 
\begin{eqnarray}
\frac{\kappa}{2}\,S(s)=\tau_P\, Im[g_P(s)/g_0(s)] + \tau_f
\,Im[g_f(s)/g_0(s)] +\tau_\omega \,Im[g_\omega(s)/g_0(s)] \\ \newline+ S(s)
\{\tau_P\, Re[g_P(s)/g_0(s)] +
\tau_f\, Re[g_f(s)/g_0(s)] + \tau_\omega \,Re[g_\omega(s)/g_0(s)]\}, 
\end{eqnarray}
evaluated at $s=s_1$ to provide a third independent equation which can be
used with Eqs.(10) and (11) to determine $\tau_P, \tau_f $ and
$\tau_\omega$. One should note that
\begin{eqnarray} 
Re[g_P(s)/g_0(s)] + 
Re[g_f(s)/g_0(s)] + Re[g_\omega(s)/g_0(s)]&=&1, \\ \newline
Im[g_P(s)/g_0(s)] + Im[g_f(s)/g_0(s)]
+Im[g_\omega(s)/g_0(s)]&=&0.
\end{eqnarray}
Then one can solve Eq.(10) for $\tau_P$ in terms of the differences
$\Delta_f=\tau_f-\tau_P$ and $\Delta_\omega=\tau_\omega-\tau_P$. The two
remaining equations can then be solved for $\Delta_f$ and $\Delta_\omega$. It
is clear that this method is not limited to the specific model chosen here
and one could carry through the exercise even if more terms are needed in the
Regge fit by using measurements at additional energies. The spin-flip
factors for the different Regge poles are interesting quantities to know
within the context of any given model.

\section{}
Here we would like to be a little more adventuresome and see if we can
determine the energy dependence from {\em existing } data by making an
additional, plausible assumption; namely, it is easy to see that if
$\tau_f=\tau_\omega$ that the previously described process can be carried
through using measurements {\em at only one energy}. This assumption is not
completely arbitrary; it follows from the assumption of exchange degeneracy
for Regge couplings and trajectories and has been much used in the past
\cite{Berger}. It is, however, on shaky foundations and not always
successful phenomenologically
\cite{Irving}. We use it here without further apologies because we need it,
and, anyhow, we will soon know if it it true or not. We hope at the least
that the results to be given below will be of some use in giving some
realistic possiblilities. Obviously, they should not be used for polarimetry
without some confirmation.
We will proceed in the following way: we will first determine the real and
imaginary parts of $\tau_{pC}(42)$ using the data from E950 reported at Spin
2000 \cite{Spin 2000}, with its error ellipse. This will then be converted
into values for $\tau_P$ and $\tau_R=\tau_f=\tau_{\omega}$ , with errors.
Subsequently, this will be used to calculate the analyzing power at the
higher RHIC energies with lab momentum
$p_L=100 \,{\rm GeV/c}$ and $p_L=250 \, {\rm GeV/c}$ for proton on fixed
carbon target. 

We use a modification of the method given in the paper of Buttimore et al
\cite{Buttimore} to extract the value of $\tau_{pC}$ from the data. Starting
from the formula of \cite{K&T}
\newpage
\begin{eqnarray}
\frac{16\,\pi}{(\sigma^{pC}_{tot})^2}\,
\frac{d\,\sigma_{pC}}{d\,t}\,A^{pC}_N(t,\tau) =
\frac{\sqrt{-t}}{m_N}\,
F_C^h(t)\,\biggl\{F_C^{em}(t)\,\frac{t_c}{t}\,
\Bigl[(\kappa-2 Re[\,\tau(s)])(1-\delta_{pC}\,\rho_{pC})\nonumber \\-2
Im[\,\tau(s)](\rho_{pC}+ \delta_{pC})
\Bigr]
+2\,F_C^h(t)\Bigl(Im[\,\tau(s)] (1 +\rho_{pC}^2) \Bigr)\biggr\}\ ,
\end{eqnarray}
where $F_C^{em}(t)$ is the electromagnetic form-factor and $F_C^h(t)$ is
the hadronic form-factor for carbon; these are calculated in \cite{K&T}.
$t_c=-8 \pi Z\alpha/\sigma^{pC}_{tot}$ and 
$\rho_{pC}$ denotes the ratio of real to imaginary parts of the $pC$
amplitude (It depends on $t$ even if $\rho$ for $pp$ does not; it is also
calculated in
\cite{K&T}.) $\delta_{pC}$ denotes the Bethe phase \cite{K&Ta}; it will not
be important in this calculation. Note that $d\sigma_{pC}/dt$
has an implicit dependence on $\tau$ but it is insignificant.

We now propose fitting the $t$-dependence of
\begin{eqnarray}
\frac{A_N(t,\tau)}{A_N(t,0)} &=&(1 - \frac{2}{\kappa}Re[\,\tau(s)]) \\
\nonumber &+& \frac{2}{\kappa} Im[\,\tau(s)] \Bigl((1 +\rho_{pC}^2(t))
(t/t_c) (F_C^h(t)/F_C^{em}(t)) -
\rho_{pC}(t)
\Bigr)
\end{eqnarray}
to the data for the measured analyzing power divided by the pure CNI
analyzing power. In this way we will determine $\tau$. (Of course, this
part of the exercise is not special to the method we are proposing, but we
need the results for input to our determination of $\tau_P$ and $\tau_R$.) We
use the functions
$F_C^h(t)$, $F_C^{em}(t)$ and
$\rho_{pC}(t)$ as calculated in
\cite{K&T}.  The coefficient of $Im[\,\tau(s)]$ is a smoothly varying
function of $t$, nearly linear, the familiar bump structure being due to the
rapid variation of the denominator in $A_N$. We will use a two parameter
linear regression to determine the best values for
$1-2\,Re[\,\tau(s)]/\kappa$ and $2\, Im[\,\tau(s)]/\kappa $. In Fig.1 we show
the result for the best fit obtained, along with the data and
the $1\sigma$ band. The $\chi^2$ for the fit is quite good, 2.1
for 4 degrees of freedom. Alternatively, the usual fit to $A_N$
itself is shown in Fig. 2.

\begin{figure}[thb]
\centerline{\epsfbox{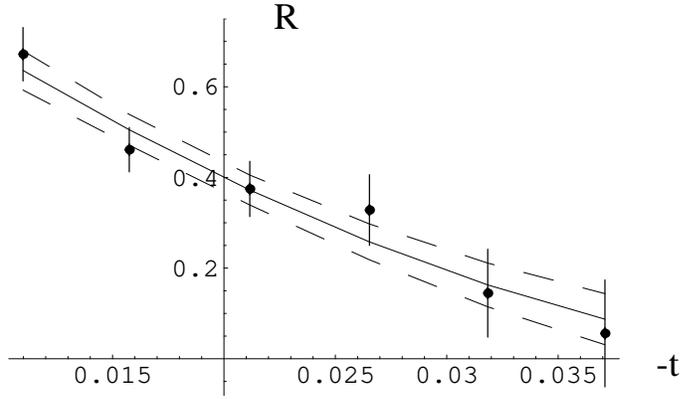}}
\medskip
\caption{\sl The best fit to the ratio of the measured analyzing
power as given at the Spin 2000 conference to the pure CNI analyzing power.
The 1 $\sigma$ confidence band is shown by dashed curves.}
\end{figure}
\begin{figure}[thb]
\centerline{\epsfbox{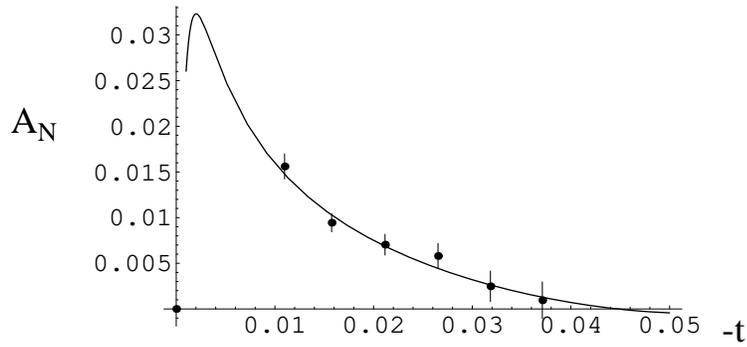}}
\medskip
\caption{\sl The best fit to the measured analyzing
power as given at the Spin 2000 conference.}
\label{
comparison}
\end{figure}

The best fit gives $Re[\,\tau(42)]=0.010$ and $Im[\,\tau(42)]=-0.038$; the
errors on these two values are considerable and correlated, so we show the
error ellipse in Fig.\,3.
\begin{figure}[h]
\centerline{\epsfbox{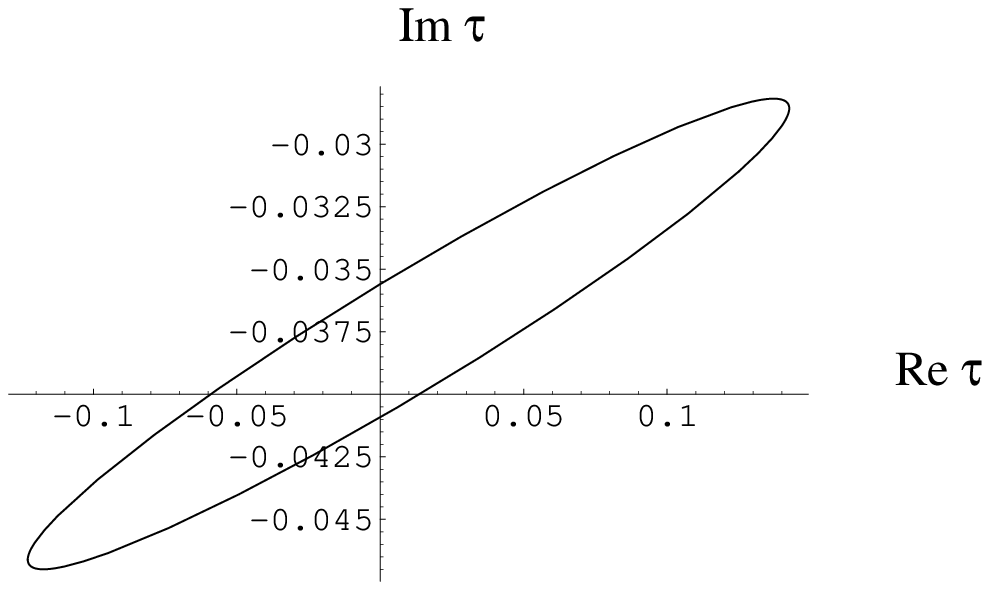}}
\medskip
\caption{\sl The 68.3\% confidence level ellipse for the linear regression of
the E950 data. }
\label{
ellipse}.
\end{figure}
We see from this that $Re[\,\tau]$ is very uncertain; it is 
consistent with zero but could be bigger than 0.1 in magnitude with either
sign.  $Im[\,\tau]$  is much better determined, to within about 15\% of its
value; it is definitely not zero, and is negative. This better determination
depends on the high sensitivity to $Im[\,\tau]$ of the prediction
of the analyzing power at higher $|t|$.
\clearpage
Now we use Eq.(10) and Eq.(11) to solve for $\tau_P$ and
$\tau_R=\tau_f=\tau_\omega$, by our assumption. The results for the central
values are $\tau_P=0.064$ and $\tau_R=-0.078$. The error ellipse is shown in
Fig.4; the error on $\tau_R-\tau_P$ is relatively small because it is
determined by $Im[\,\tau(42)]$. The error on $\tau_P$ itself is quite large;
it is consistent with asymptotic spin dependence vanishing or being as large
as 15\%. 

\begin{figure}[h]
\centerline{\epsfbox{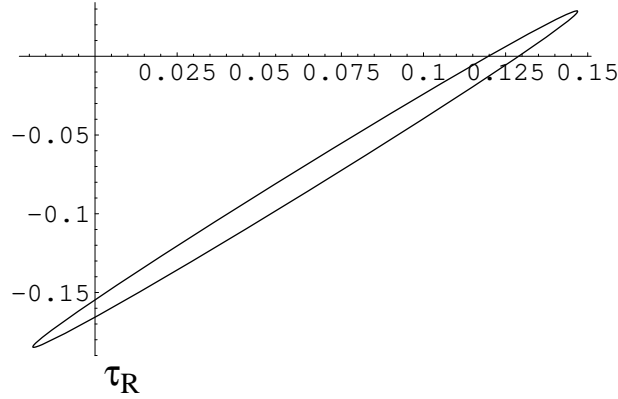}}
\medskip
\caption{\sl The 68.3\% confidence level ellipse for the Regge spin-flip
couplings $\tau_P$ and $\tau_R$
.}
\label{
Reggeellipse}
\end{figure}

It is more interesting now to see the implication for the near term
RHIC measurements with 100 GeV and 250 GeV proton beams on a fixed target. In
this range both $Re[\,\tau(s)]$ and $Im[\,\tau(s)]$ vary, but not by enormous
amounts. This is shown in Fig.\,5.
\begin{figure}[thb]
\centerline{\epsfbox{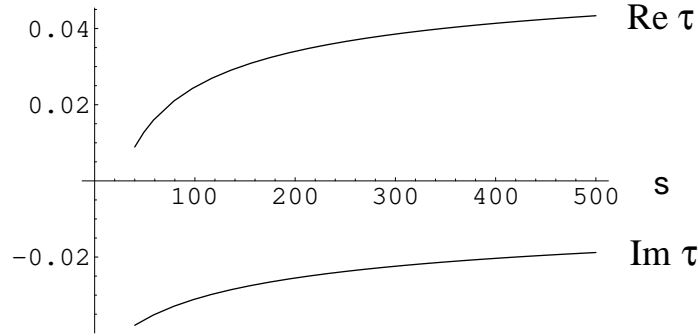}}
\medskip
\caption{\sl The energy dependence of  the real and  the imaginary
parts of the pC spin-flip factor over the RHIC energy range.}
\label{
tau energy depe.}
\end{figure}
The ratios to pure CNI, as in Fig. 1, are given in Fig.\,6 and the
corresponding plots of $A_N$ in Fig.\,7.
\begin{figure}[thb]
\centerline{\epsfbox{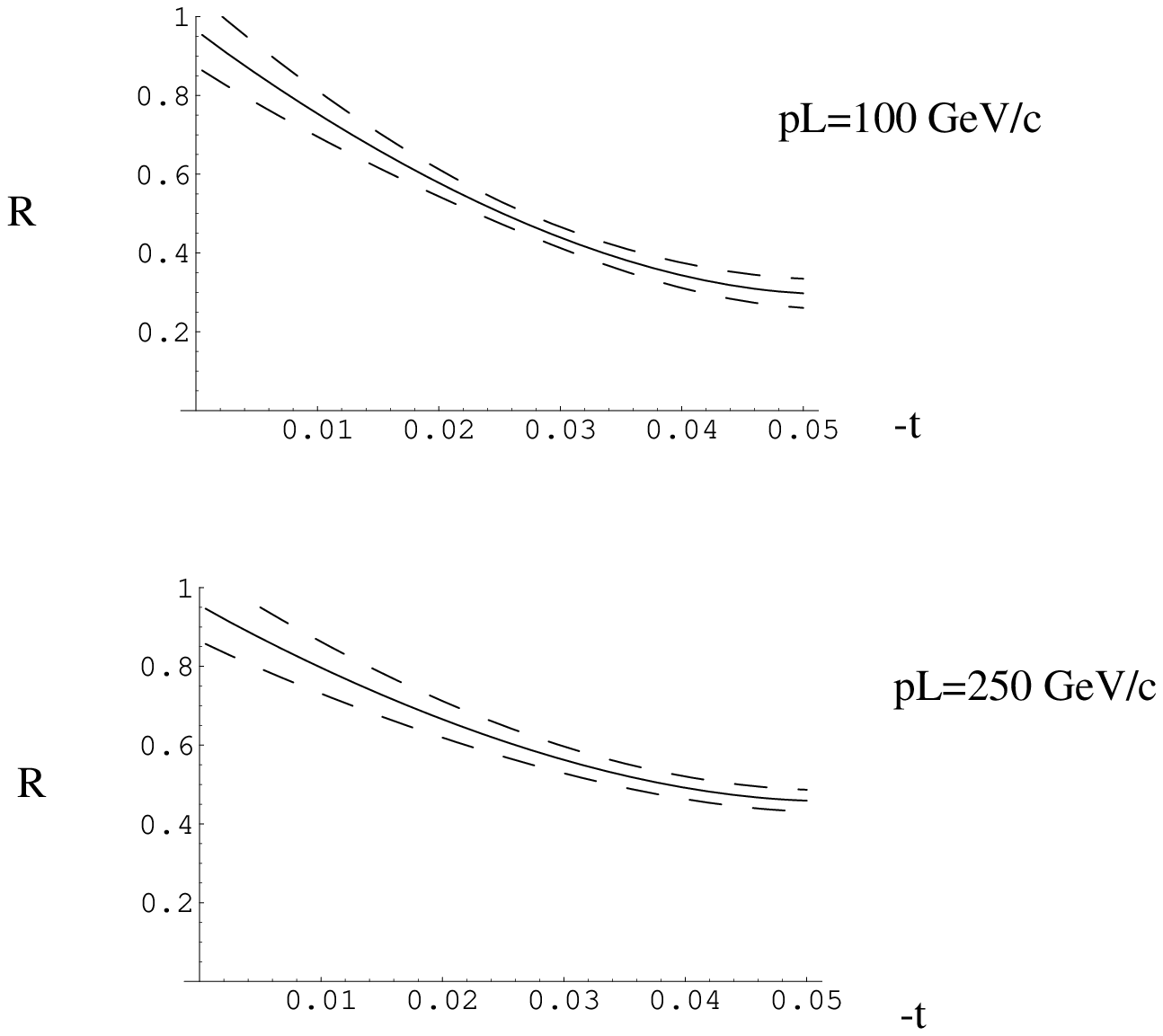}}
\medskip
\caption{\sl The ratio of the predicted analyzing power to pure CNI at
$p_L=100\,
GeV/c$ and $p_L=250 \,GeV/c$ compared to the best fit curve at $p_L=21.7\,
GeV/c$. The 1 $\sigma$ bands are also shown.}
\label{
ratios}
\end{figure}
The asymptotic values will be eventually $Re[\,\tau(s)]=0.064$ while
$Im[\,\tau(s)]$ will vanish; we see that we are a long way from that
situation. 
\begin{figure}[thb]
\centerline{\epsfbox{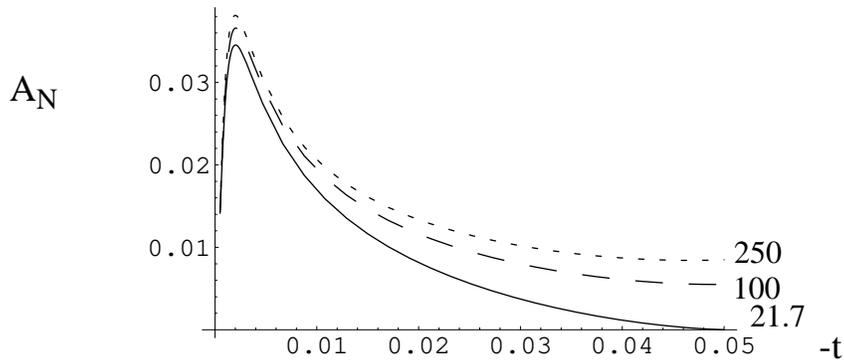}}
\medskip
\caption{\sl The predicted analyzing power  at
$p_L=100\,
GeV/c$ and $p_L=250 \,GeV/c$ compared to the best fit curve at $p_L=21.7\,
GeV/c$. }
\end{figure}

\clearpage

\section{}
Does this analysis teach us anything about the $pp$ analyzing power in the
RHIC energy range? As pointed out at the beginning, the $I=1$ contribution
is missing from $pC$ scattering; although it is known that, even as low as
the E950 energy, the non-flip $I=1$ is small, the $I=1$ flip is very likely
not to be small. It is well-known from Regge fits to $\pi p$ scattering that
the $\rho$ spin-flip coupling is very large. Indeed, from the global fits
done by Irving and Worden \cite{Irving} the $I=1$ spin-flip Regge couplings
are taken to be nearly an order of magnitude larger than the $I=1$ non-flip
couplings. These are used in the analysis of Berger et al \cite{Berger}.

We can determine the error ellipse for the E704 data and compare it with the
ellipse projected to $s=400$ for the $I=0$ piece from the $pC$ fits above.
This is shown in Fig.8. Even though the E704 error ellipse is enormous, the
central values, especially for $Im[\,\tau]$, are so different that there is
no overlap of the two regions. The same remains true if we enlarge the
ellipses to $90\%$ confidence level.

\begin{figure}[thb]
\centerline{\epsfbox{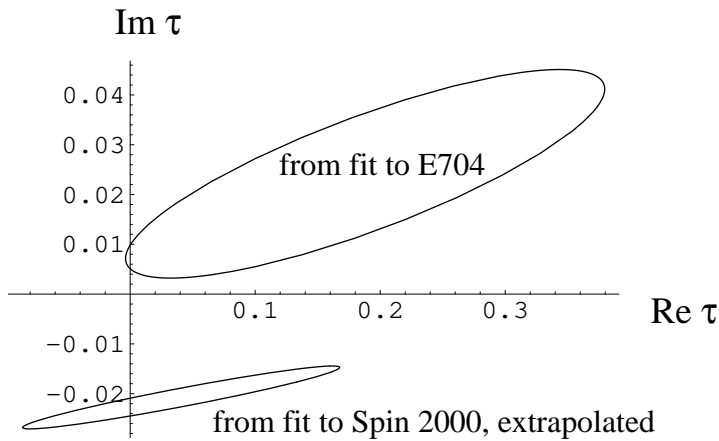}}
\medskip
\caption{\sl Ellipse from fitting E704 data compared to pC ellipse
extrapolated to the same energy.}
\label{
pp comp.}
\end{figure}

The data is too uncertain to try to do a serious fit to determine the $I=1$
piece from this difference. However, if one assumes that the Pomeron
coupling is well determined from the $pC$ data because it is $I=0$, then we
can try the same three Regge pole model and determine the size of the $C=+1$
and $C=-1$ Regge coupling (summing both iso-spins) and find that for
$C=+1$, $\tau_+=0.53$ and for $C=-1$, $\tau_-=-1.03$. These are to be
compared with the $I=0$ Regge coupling determined in Section 3:
$\tau_R=-0.078$, evidently about an order of magnitude smaller, consistent
with ancient expectations.

\section{}
The results presented here must be considered to be preliminary. The
numerical work has been checked by me several times, but not by anyone else.
This all needs to be done more carefully when more data becomes available,
which is expected to be soon. The parameters found are quite small but for
larger $|t|$ for carbon (or any heavier nuclear target) the modification
from pure CNI is large. Since the analysis here uses only the statistical
errors, it seems likely that, in the small $t$-region, near the peak,
the experimental uncertainty will be larger than the effects calculated here,
which are less than $\approx 10\%$. Of course, this would be a useful
result and is probably not very sensitive to the details of the model.

The Regge model used in the first section is fairly standard, but as with
all Regge models, is pure phenomenology and the magnitude of the parameters
are not calculable. As soon as data becomes available at a higher energy,
the determinations described there can be carried through and, possibly,
checked for use at higher energy. The model described in the second
section is more speculative, requiring the assumption that
$\tau_f=\tau_\omega$ ; this is unknown but is not unreasonable, and it allows
us to make quantitative predictions for higher energy, just based on the
E950 data. All of the quantitative results in Section 2 and Section 3 depend
on it. We will soon know if it is wrong, but in the short term it may
provide some guidance regarding reasonable expectations for energy
dependence. The values of the spin-flip Regge couplings hold intrinsic
interest for subsequent phenomenology and, possibly, for eventual
understanding of the spin-dependence of Regge couplings.

I would like to thank Nigel Buttimore and Boris Kopeliovich for very helpful
comments on this note.

\end{document}